\documentstyle[12pt]{article}

\def\RR{{{\rm l}\kern-.17em{\rm R}}}
\newcommand\bfab{\em}

\newcommand\xh{{\hat{x}}}

\newcommand\Dub{{\overline{\Delta}_N}}

\newtheorem{thm}{Theorem} 
\newtheorem{lemma}[thm]{Lemma} 
\newtheorem{cor}[thm]{Corollary} 
\newcommand{\EQ}{\begin{equation}}
\newcommand{\EN}{\end{equation}}
\newcommand{\BEQ}{\begin{equation}}
\newcommand{\NEQ}{\end{equation}}
\newcommand{\nn}{\nonumber}

\def\ni{\noindent}

\def\eopp{{$\Box$}}      

\def\Ebf{{\bf E}\,}
\def\htl{\tilde{h}\,}

\def\Qt{\tilde{Q}\,}

\def\fl{{f^{(\ell)}}}
\def\flp{{f^{(\ell+1)}}}

\def\fhl{\hat{f}^{(\ell)}}

\def\fhlN{\hat{f}_N^{(\ell)}}
\def\fhN{\hat{f}_N^{}}
\def\fhlpN{\hat{f}_N^{(\ell+1)}}
\def\eps{{\epsilon}}

\def\part{\partial}

\date{ }

\begin{document}
\bibliographystyle{plain}

\title{
Improved Asymptotics for  Zeros 
of Kernel Estimates via
a Reformulation of the  Leadbetter-Cryer Integral
\footnote{We  thank the referee for useful comments. 
Research funded by the U.S. Department of Energy. 
}
}

\author{Kurt S.~Riedel \\
Courant Institute of Mathematical Sciences \\
New York University \\
New York, New York 10012-1185}

{\vskip-.25in}
\maketitle
\begin{abstract} 
The expected number of false inflection points of kernel smoothers
is evaluated. To obtain the small noise limit, we use a 
reformulation of the Leadbetter-Cryer integral for the expected
number of zero crossings of a differentiable Gaussian process. 
\end{abstract}

\section{Convergence of Kernel Smoothers}
\label{WLEsect}

For many applications of nonparametric function estimation, obtaining
the correct shape of the unknown function is of importance. 
A consequence of Mammen et al. (1992, 1995) is that kernel smoothers
have a nonvanishing probability of having spurious inflection points
if the smoothing level is chosen to minimize the mean integrated square 
error (MISE).  In Riedel (1996), we propose a two-stage estimator
where the number and location of the change points is estimated using
strong smoothing. 

In this letter, we evaluate the probability of obtaining spurious inflection
points for kernel smoothers in the small noise/heavy smoothing limit.
The proofs are based on powerful and seldom used techniques:
Koksma's theorem and the  Leadbetter-Cryer integral for the expected
number of zeros of a differentiable Gaussian process.


We consider a sequence of kernel smoother estimates, $\fhN(t)$,
of  $f(t)$, 
and examine
the convergence of the estimate as the number of measurements, $N$, increases.
We believe that our results are  slightly
stronger than previous theorems on kernel smoothers 
(Gasser \& M\"uller 1984). 
For each $N$,
the measurements occur at $\{ t_i^N, \ i= 1 \ldots N \}$.
We suppress the superscript, $N$,
on the measurement locations $t_i \equiv t_i^N$. 
We define the empirical
distribution of measurements, $F_N(t)= \sum_{t_i \le t} 1/N$, 
and let $F(t)$ be its limiting distribution.

\noindent
{\bf Assumption A} 
{\em Consider the sequence of
estimation problems: $y_i^N = f(t_i^N) + \eps_i^N$,
where the $\eps_i^N$ are zero mean random variables and 
${\bf Cov}[\eps_i^N,\eps_j^N] =\sigma^2 \delta_{i,j}$.
Assume that the distribution of  measurement locations 
converges in the sup norm: $D^*_N \equiv\ \sup_t \{|F_N(t)- F(t)|\} 
\rightarrow 0$,
where $0<c_F<F'(t)<C_F$. 
}


The star-discrepancy, $D^*_N \equiv\ \sup_t \{ F_N(t) -F(t) \}$, 
is useful because it  measures how closely 
a discrete sum over an arbitrarily placed set of points approximates
an integral. (See Theorem \ref{Kokcor}.) 
For regularly spaced points, $F(t_i) =(i+.5)/N$
and $D_N^* \sim 1/N$, while for randomly spaced points, 
$D_N^* \sim \sqrt{\ln[\ln[N]]/N}$ by the Glivenko-Cantelli Theorem.

We consider kernel estimates of the form: 
\begin{equation}\label{K1}
\fhl(t) = \frac{1}{Nh_N^{\ell+1}}
\sum_i^N \frac{y_iw_i}{F'(t_i)} \kappa^{(\ell)}({t-t_i \over h_N}) \ ,
\end{equation}
where $h_N$ is the kernel halfwidth and  $\{w_i\}$ are weights.
We need convergence results for kernel estimators, $\fhlN(t)$,
of  $f^{(\ell)}(t)$. 
Our hypotheses are stated in terms of the star discrepancy
while previous results impose stronger/redundant conditions.
We define 
$\sigma^2_{N}(t) = {\bf Var}[\fhlN(t)]$,
$\xi^2_{N}(t) = {\bf Var}[\fhlpN(t)]$,
$\mu^2_{N}(t) = {\bf Corr}[\fhlN(t),\fhlpN(t)]$.
We now evaluate the limiting quantities for a class of kernel smoothers.
We use the notation ${\cal O_R}(\cdot)$ to denote a  size of ${\cal O}(\cdot)$
relative to the main term: 
${\cal O_R}(\cdot)=\times \ [1+ {\cal O}(\cdot)]$.
We denote $C^{\ell}$ as the set of $\ell$ times continuously differentiable
function, $TV[0,1]$ as the function of bound variation with the total
variation norm, $\|\cdot\|_{TV}$.
We define $\|f\|_{bv}$ to be the sum of the $L_{\infty}$ and total
variation norms of $f$ and define $\|f\|$ to be the $L_{2}$ norm.

\begin{thm}[Generalized Gasser-M\"uller (1984)     
]\label{KS}
Let $f(t) \in C^{\ell+1}[0,1]\cap  TV[0,1]$ 
and consider a sequence of estimation
problems satisfying Assumption A. Let
$\fhlN(t)$ be a kernel smoother estimate as given in (\ref{K1}),
where the halfwidth, $h_N$, 
and the weights, $\{w_i\}$,
satisfy $|w_i-1| \sim {\cal O}(D_N^*/h_N)$. 
Let the kernel, $\kappa^{(\ell+1)} \in TV[-1,1]\cap C[-1,1]$,
satisfy the moment condition:
$\int_{-1}^1  \kappa(s)ds = 1$,  
and the boundary conditions:
$\kappa^{(j)}(-1) =\kappa^{(j)}(1) = 0$ for $0\le j \le \ell$.
Choose the kernel halfwidths such that $h_N\rightarrow 0$, 
and $D_N^*/h_N^{\ell +2} \rightarrow 0$; 
then 

\ni
i)
$\Ebf[\fhlN](t) \rightarrow f^{(\ell)}(t)
 + {\cal O_R}(h_N + D_N^*/h_N^{\ell +1})$,

\ni
ii) $\Ebf[\fhlpN](t) = 
\int_{-1}^1 f^{(\ell+1)}(t+hs)\kappa(-s)ds $  
$+ {\cal O}(\|f\kappa^{(\ell+1)}\|_{bv} D_N^*/h_N^{\ell +2})$,

\ni
iii) $\sigma_N^2(t)  \rightarrow \sigma^2\|\kappa^{(\ell)} \|^2
/ (N F'(t) h_N^{2\ell +1} ) +\ 
{\cal O_R}(h_N + D^*_N/h_N)$, \ 

\ni
iv) 
$\xi_N^2(s) \rightarrow  \sigma^2$
$\|\kappa^{(\ell+1)} \|^2_{}$ $ / (NF'(s)h_N^{2\ell +3} ) 
+\ {\cal O_R}( h_N+ D^*_N/h_N)$, and  \ 

\ni v) 
$\mu^2_N(t) \rightarrow {\cal O}(h_N +\  D_N^*/h_N)$ 

\ni
uniformly 
in the interval, $[h_N,1-h_N]$. 
\end{thm}

Our proof of Theorem \ref{KS} is based on Koksma's Theorem which
bounds the difference between integrals and discrete sum approximates: 

\begin{thm} [Generalized Koksma Niederieter (1992) 
] 
\label{Kokcor}
Let $g$ be a bounded function of bounded variation, $\|g\|_{TV}$, on $[0,1]$:
$g \in TV[0,1] \cap  L_{\infty}[0,1]$.
Let the star discrepancy be measured by a distribution, $F(t)\in C^{1}[0,1]$ 
with $0<c_F<F'(t)<C_F$. 
If the discrete sum weights, $\{w_i, i= 1, \ldots N \}$, 
satisfy $|w_i -1| \le C D_N^*$, then
\BEQ
\left| \int_0^1 g(t)dF(t) - \frac{1}{N}\sum_{i=1}^N g(t_i)w_i 
\right| \le \left[ \|g\|_{TV} + C \|g\|_{\infty} \right] D_N^*   \ .
\NEQ
\end{thm}

In our version of Koksma's Theorem, we have added two new effects:
a nonuniform weighting,  $\{w_i, i= 1, \ldots N \}$, and a nonuniform
distribution of points, $dF$.
The total variation of $g(t(F))$ with respect to $dF$ is equal to
the total variation of $g(t)$ with respect to $dt$. 
Theorem \ref{Kokcor} follows from  Koksma's Theorem
by a change of variables.


\noindent
{\bfab Proof of Theorem \ref{KS}.}  
We rescale: $s_i = (t_i-t)/h_N$ and apply Koksma's theorem
to $f(t+hs) \kappa^{(\ell)}(-s) \in TV_s[-1,1]$.
The contribution of the weights, $w_i$, is  
${\cal O_R}( D^*_N/Nh_N^{\ell+1})$. 
Thus $\Ebf[\fhlN](t) = \int_{-1}^1 f^{(\ell)}(t+hs)
\kappa(-s)ds $ 
$+ {\cal O}(\|f\kappa^{(\ell)}\|_{bv} D_N^*/h_N^{\ell +1})$.
Since ${|\kappa^{(\ell+1)} (-s)|^2}/{F'(t+h_Ns)}$ is in $TV[-1,1]$,
the variance satisfies 
$$
\xi_N^2(t)  =
\frac{\sigma^2}{ N h_N^{2\ell +1}} 
\int \frac{|\kappa^{(\ell)} (-s)|^2}{F'(t+h_Ns)} ds
+\ 
 {\cal O_R}( D_N^*/h_N) \ .
$$
The result follows from expanding $F'(t)$ in $h_N$. 
\eopp

Theorem \ref{KS} is one of two ingredients which we need to bound the
expected number of change points of $\fhlN(t)$. Section \ref{LCSec}
presents the second ingredient.

\section{Asymptotics of Zero Crossings}\label{LCSec}

The Leadbetter-Cryer (L-C) expression 
evaluates the expected number of
zeros of a differentiable  Gaussian process, $Z(t)$, in terms 
of a time history integral involving the first and second moments of $Z(t)$ 
(Leadbetter and Cryer 1965).
We reexpress this integral in terms of the zeros of $E[Z(t)]$
and a remainder term.
This alternative expression is particularly useful
in the small noise limit  when  one desires an asymptotic 
evaluation of the number of noise induced zero crossings.


\begin{thm}[Leadbetter \& Cryer (1965), 
Cram\'er \& Leadbetter, 1967, Sec.~13.2]
\label{CLthm}
Let  $Z(t)$ be a pathwise continuously differentiable Gaussian  process 
in the time interval [0,T]. Denote $m(s) ={\bf E}[Z(s)]$,
$\Gamma(s,t) = {\bf Cov}[Z(s),Z(t)]$,
$\sigma^2(s) ={\bf Var}[Z(s)]=\Gamma(s,s)$, $\xi^2(s) ={\bf Var}[Z'(s)]$,
$\mu(s) ={\bf Corr}[Z(s)Z'(s)]$. 
Let $N_z$ be the number of zero crossings of $Z(t)$.
If $m(t)$ is continuously differentiable, $\Gamma(s,t)$ has mixed
second derivatives that are continuous at $t=s$ and $\mu(s) \ne 1$ at any
point $s \in [0,T]$,
then 
\begin{equation}\label{CL} 
{\bf E}[N_z] = \int_0^T \frac{\xi(s)\gamma(s)}{\sigma(s)}
\phi\left(\frac{m(s)}{\sigma(s)}\right) Q(\eta(s))ds  \ , 
\end{equation}
where
$Q(z) \equiv 2\phi(z)+ z[2\Phi(z) -1]$,
 $\gamma(s)^2 = 1 - \mu(s)^2$,
 $\eta(s) = 
\frac{m'(s) - \xi(s)\mu(s)m(s)/\sigma(s)}{\xi(s)\gamma(s)}$.
\end{thm}

By decomposing (\ref{CL}) into two pieces, we derive the following bounds:

\begin{thm}[Alternate form] \label{CLthm1}
Let the hypotheses of Theorem \ref{CLthm} hold and
define $M(t)$ 
$\equiv \ m(t)/ \sigma(t)$.
Let $|M(t)|$ have $N_z^o$ zeros,
$L_{mx}$ relative maxima, $M_j$, $j= 1 \ldots L_{mx}$
and  $L_{mn}$ nonzero relative minima, $m_{j}\ne 0$, $j= 1 \ldots L_{mn}$,
where $M(0)$ and $M(T)$ are counted as relative  extrema.
Let $\nu_j$ equal 1 if $m_j$ occurs at  $0$ or $T$
and $\nu_j=2$ otherwise.  Define $\hat{\nu}_j$ similarly for
the $M_j$.
Equation (\ref{CL}) can be rewritten as 
\begin{equation}\label{CL1} 
{\bf E}[N_z] - N_z^0\ = \sum_{j=1}^{L_{mn}}\nu_j \Phi(-m_j)\ 
- \sum_{j=1}^{L_{mx}}\hat{\nu}_j \Phi(-M_j)\ +
 \int_0^T \frac{\xi(s)\gamma(s)}{\sigma(s)}
\phi\left(\frac{m(s)}{\sigma(s)}\right) \Qt(\eta(s))ds  \ , 
\end{equation}
where
$\Qt(z) \equiv 2 \int_{|z|}^{\infty} \phi(s')[s'-|z|]ds'$.
\end{thm}

\noindent
{\bfab Proof.} Write $Q(z) = |z| + \Qt(z)$. The  first term in (\ref{CL})
equals $-\int \Phi'(M) |M'(t)|dt$.  Integrating this term yields the weighted
sum of the relative extrema of $\Phi(-|M|(t))$. We decompose this sum into
$N^0_z$ zeros of $|M|(t)$ plus the additional relative extrema:
$\sum_{j=1}^{L_{mn}}\nu_j \Phi(-m_j)\ -
\sum_{j=1}^{L_{mx}}\hat{\nu}_j \Phi(M_j)\ $. 
\eopp

We are unaware of any previous derivation of Theorem \ref{CLthm1}.
The second term on the right hand side of (\ref{CL1}) 
corresponds to the probability that $Z(t)$ lacks a
zero of $m(t)$ while the the first and third terms correspond to extra zeros.
Note that $\Qt(z) \le \phi(z) \le 1/\sqrt{2\pi}$.

\begin{cor} \label{Csimple}
Under the hypotheses of Theorems \ref{CLthm}  \&  \ref{CLthm1},
let $\{(x_k,w_k),k=1\ldots K\}$ be chosen  such that 
$|t -x_k |$ 
$\le w_k$ implies that  $m'(t) >0$ and
${|M(t)| \ge c m'(x_k)|t-x_k|/\sigma(x_k) }$, where 
$c$ is a fixed number, $0<c<1$. 
Define $\Psi_k \equiv\ \sup_{|s -x_k|  \le w_k}
\{\Qt(s)\xi(s)\gamma(s)\sigma(x_k)/\sigma(s) \}$,
$C = sup_{t}\{ \xi(t)/\sigma(t) \}$, and 
$m_o \equiv\ \inf \{|M(s)|\ {\rm for}\ s\ {\rm such\ that\ } 
|s -x_k|  \ge w_k,\ k= 1\ldots K  \} $. 
 The expected number of zeros of the Gaussian process, $Z(t)$, satisfies
\begin{equation}\label{CLbnd} 
{\bf E}[N_z] - N_z^o \ \le \
\sum_{k=1}^{K} \frac{\Psi_k}{cm'(x_k)}  + 
{\cal O}\left((CT+2L_{mn} )\phi(m_o) \right)    
\ .
\end{equation}
\end{cor}

\noindent
{\bfab Proof.} The first term in (\ref{CLbnd}) arises from replacing
$\Psi\int_{x_k-w_k}^{x_k+w_k}\phi\left(\frac{m(s)}{\sigma(s)}\right)ds$
by $\Psi\int_{-\infty}^{+\infty}
\phi\left(\frac{cm'(x_k)s}{\sigma(x_k)}\right)ds$ and integrating.
\eopp

A sufficient additional condition for the existence of a set of $(x_k,w_k)$
satisfying Corollary \ref{Csimple} is that $m(s)$ vanishes only at
a finite number of points, $\{x_k\}$, and at these points, $m'(x_k)\ne 0$.
Let $\delta$ be a small parameter related to the weakness of the noise 
amplitude. In many cases, the $\{w_k\}$ can be chosen to be powers of 
$\delta$ and the upper bound of (\ref{CLbnd}) reduces to 
\begin{equation}\label{CLbnd1} 
{\bf E}[N_z] - N_z^o \ \le \
\sum_{k=1}^{K} \frac{\Qt(x_k)\xi(x_k)\gamma(x_k)}{cm'(x_k)}   
\left[1 +  { o}(1) \right] \ .    
\end{equation}
In contrast, a similar naive expansion of the original integral (\ref{CL})
yields the asymptotic expression:
\begin{equation}\label{CLbnd2} 
{\bf E}[N_z] - N_z^o \ \le \   N_z^o  { o}(1) \ + \
\sum_{k=1}^{K} \frac{\Qt(x_k)\xi(x_k)\gamma(x_k)}{cm'(x_k)} 
\left[1 +  { o}(1) \right] \ .    
\end{equation}
The advantage of (\ref{CLbnd1}) over (\ref{CLbnd2}) is that the
remainder term, $ N_z^o { o}(1)$, has been integrated away. 




\section{Number of false change points} \label{CLsect}

We now consider sequences of kernel 
estimates of $f^{(\ell)}(t)$,
and examine the number 
of false $\ell$-change points.  We restrict to independent
$Gaussian$ errors: 
$\eps_i \sim N(0,\sigma^2)$. 
Thus, $\fhlN(t)$ is a Gaussian process.
Mammen et al.\ (1992,1995) consider the statistics of  
change point estimation 
for kernel estimation of a probability density. We present the analogous
result for regression function estimation. In both cases, the analysis
is based on the Leadbetter-Cryer formula for zero crossings.
The following assumption rules out nongeneric cases:

\

\noindent
{\bf Assumption B} {\em
Let $f(t) \in C^{\ell+1}[0,1]$ have $K$ $\ell$-change points,
$\{x_1, \ldots x_K\}$, with $f^{(\ell )}(x_k)=0$, $f^{(\ell +1)}(x_k)$
$\ne 0 $,
$f^{(\ell )}(0) \ne 0 $ and $f^{(\ell)}(1) \ne 0 $.
Consider a sequence of estimation problems with
independent, normally distributed measurement 
errors, $\eps_i^N$, with variance $\sigma^2$.
Let $\fhlN(t)$ be a sequence of  
kernel estimates of $f^{(\ell)}$, 
on the sequence of intervals, $[\delta_N,1-\delta_N]$.
}

Gasser and M\"uller (1984) evaluate the variance of a change point
estimate:
${\bf Var}[\xh_k- x_k] \approx
{\sigma}_{\rm if}^2({x}_k) \equiv\
{ {\bf Var}[\fhlN(x_k)] / |{f^{(\ell+1)}}({x}_k)|^2 }$ . 
The following theorem bounds the tail of the empirical change point 
distribution $|\xh_k-x_k| >>  {\sigma}_{\rm if}$. 
By using the L-C integral, we require weaker conditions than
the hypotheses of Gasser and M\"uller (1984). 

\begin{thm}\label{KZ}
Let 
Assumption B hold and consider a sequence of kernel estimators, 
$\fhlN(t)$, that  satisfy the hypotheses of Lemma \ref{KS}. 
Choose kernel halfwidths, $h_N$, and uncertainty intervals, $w_N$,
such that
$h_N/w_N \rightarrow 0$,  $w_N \rightarrow 0$,
$w_{N,k}^2 N h_N^{2\ell +1} \ge 1$.
The probability, $p_N(w_N)$, that 
$\hat{f}^{(\ell )}_N$ has a false change point outside of a width of 
$w_N$ from the actual $(\ell+1)$-change points satisfies 
\BEQ \label{Kbnd1}
p_N(w_N) \le \
\sum_{k=1}^K {\cal O}
\left( \frac{\sigma_{if}(x_k)}{h_N} 
\exp\left({-w_N^2 \over 2\sigma_{if}^2(x_k)} \right)\ \right) \ ,
\NEQ
where $\sigma_{if}^2(x_k) \rightarrow 
\sigma^2 \|\kappa^{(\ell)}\|^2 \left/ |\flp(x_k)|^2NF'(x_k)h_N^{2\ell +1}
\right.$ 
on the interval $[h_N,1-h_N]$.
\end{thm}

\noindent
{\bfab Proof.} 
Lemma \ref{KS} shows that 
$\xi_N(t)/ \sigma_{N}(t) \rightarrow {\cal O}(h_N^{-1})$.
Within a neighborhood of $\sqrt{w_N}$ of $x_k$,
$ {\bf E} [\fhlN(t)] = f^{(\ell+1)}(x_k)(t-x_k) \ 
+ {\cal O}(\sqrt{w_N} + D^*_N/h_N^{\ell +1})$.
Define $b_N  = \inf\{  |f^{(\ell)}(t)|\ { such\ that}\ t \notin 
\cup_{k=1}^K (x_k- \sqrt{w_N}, x_k + \sqrt{w_N})\}$.
Note that $b_N \ge C \sqrt{w_N}$ asymptotically and
the integral of (\ref{CL}) outside of 
$\cup_{k=1}^K (x_k- \sqrt{w_N}, x_k + \sqrt{w_N})$
is bound by
$\exp(-c{w_N}/{\sigma_{N}^2})<<\exp(-{w_N^{2}}/{2\sigma_{if}^2}(x_k))$.
Integrating the ${\cal O}(1)$ integrand bound, 
$\exp\left(-|f^{(\ell+1)}(x_k)|^2|t-x_k|^2/2\sigma_N^2(x_k)\right)/h_N$,
over the intervals $[x_k\pm \sqrt{w_N}, x_k \pm w_N]$  yields
(\ref{Kbnd1}).
\eopp

Mammen et al.\ (1992,1995) derived the number 
of false change points 
for kernel estimation of a probability density for {\em nonvanishing}
error probabilities. We now show that there expression remains
valid as the error probability goes to zero.
Given Gaussian measurement errors,   
the sophisticated proof in  Mammen (1995) 
can be 
simplified in our case. 

\begin{thm}[Analog of Mammen et  al.\ (1992,1995)]
\label{MMFthm}
{Let 
Assumption B hold.
Consider a sequence of kernel smoother estimates $\hat{f}_N$ which 
satisfy the hypotheses of Lemma \ref{KS} with 
$\int_{-1}^1 s \kappa(s)ds$ 
$=0$. 
Let the sequence of kernel halfwidths, $h_N$, satisfy
 $D_N^* N^{1/2} h_N^{\frac{1}{2}} \rightarrow 0 $ and
$0<{\rm liminf}_N h_N N^{{1}/{(2\ell+3)}} $
$\le{ \rm limsup}_N h_N N^{{1}/{(2\ell+3)}}$ $< \infty $.
The expected number of $\ell$-change points of  $\hat{f}_N$
in the estimation region, $[h_N,1-h_N]$,
is asymptotically
\begin{equation}\label{Mam1} 
{\bf E}[\hat{K}] - K
=  \ 2\sum_{k=1}^K 
H\left(\sqrt{ { | f^{(\ell +1)}(x_k)|^2 NF'(x_k)h^{2\ell+3}
\over \sigma^2\|\kappa^{(\ell +1)}\|^2  } }
\right)  \ + o_{\cal R}(1)\
, 
\end{equation}
where 
$H(z) \equiv \phi(z)/z +\Phi(z) -1$ with $\phi$ and $\Phi$ being the 
Gaussian density. 
If $\flp(t)$ has H\"older smoothness of order $\nu$ for some $0<\nu<1$, 
and $ h_N N^{{1}/{(2\ell+3)}} \rightarrow 0 $,
then (\ref{Mam1}) remains valid provided that  
$ h_N N^{{1}/{(2\ell+3+ 2\nu)}} \rightarrow 0$.

}
\end{thm}

In Mammen (1992,1995), 
the correction in (\ref{Mam1})
is shown to be $o(1)$  if   
${ \rm limsup}_N$ $h_N N^{{1}/{(2\ell+3)}}$ $< \infty $.
We strengthen this result by showing that (\ref{Mam1}) continues
to represent the leading order asymptotics even when
$ h_N N^{{1}/{(2\ell+3)}} \rightarrow \infty $.
Our secret is to use (\ref{CL1}) instead of (\ref{CL}) because
(\ref{CL1}) has integrated out the term equal to $K$.

\noindent
{\bfab Proof of Theorem \ref{MMFthm}.} 
Theorem \ref{KZ} shows that the contribution away from the
$\ell$-change points is exponentially small for $|s-x_k| >> \sigma_N(s)$.
Lemma \ref{KS} shows that $\frac{\xi_N(s)\gamma_N(s)}{\sigma_N(s)}
\rightarrow \frac{\|\kappa^{(\ell+1)} \|}{h_N\|\kappa^{(\ell)} \|}$
and that for $|s- x_k|<<1$, 
$\eta_N(s) \rightarrow {\flp(s)}/{\sigma_N(s) }$.

Equation (\ref{Mam1}) is an approximation of (\ref{CL1}) 
using Laplace's method. To prove (\ref{Mam1}), we must show that 
${\bf E}[\fhlN(t)] = \fl(t)  + o_{\cal R}(\sigma_N)$
for $|t-x_k| \sim \sigma_N$.
Near the change point, $x_k$, 
\begin{eqnarray} \label{oh}
{\bf E}[\fhlN(t)] &=& \fl(t) + 
\int_{-1}^{1} \kappa(s)\left[\fl(t+h_Ns) - \fl(t)\right]ds
+\ {\cal O_R}(D_N^*/h_N^{\ell +1})
\nn \\
&=&  \fl(t) +\ h_N
\int_{-1}^{1} s \kappa(s)\left[\flp(t+h_N\tau_N(s)) - \flp(t)\right]ds \ ,
\end{eqnarray}
where $\tau_N(s)$ lies in $[0,s]$ by the mean value theorem.
Since $\flp(t)$ 
is continuous at $x_k$, 
for each $\delta$, there is a $\htl_N(\delta)$ such that
$|\flp(t+h_N\tau_N(s)) - \flp(t)|< \delta$
for all $t$, $t+h_N\tau_N \in [x_k-\htl_N(\delta),x_k-\htl_N(\delta)]$.
Thus ${\bf E}[\fhlN(t)] = \fl(t) +  
{\cal O_R}(\delta h_N + D_N^*/h_N^{\ell +1 })$.
Here $\delta$ may be taken arbitrarily small. 
Applying the Laplace's method yields (\ref{Mam1}) with
corrections of ${\cal O_R}\left(\exp( -\delta h_N/\sigma_{if}) -1\right)$ 
$+ {\cal O_R}\left(\exp(-D_N^*/h_N^{\ell +2}\sigma_{if})- 1\right)$.
The scaling,  $ h_N \sim N^{{-1}/{(2\ell+3)}}$, implies that the
first term is ${\cal O_R}(\delta)$.
The discrete sampling effect (the second term) requires the hypothesis that 
$D_N \sqrt{h_N N} \rightarrow 0 $ to be $o_{\cal R}(1)$.
When  $\flp(t)$ is H\"older of order $\nu$, we have the stronger bound: 
$|\flp(t+h_N\tau_N(s)) - \flp(t)|< C_t h_N^{\nu}$, and 
${\bf E}[\fhlN(t)] = \fl(t)  + {\cal O_R}(h_N^{1+\nu}+ D_N^*/h_N^{\ell +1 } )$.
The next order correction in Laplace's method is
${\cal O_R}\left(\exp( h_N^{1+ \nu}/\sigma_{if})\right)$.
This term is  $o_{\cal R}(1)$ when
$ h_N N^{{1}/{(2\ell+3+ 2\nu)}} \rightarrow 0 $.
\eopp

In Riedel (1996), we propose a two-stage nonparametric function estimator 
which 
achieves the correct shape with high probability.
In the first stage, we estimate the number and approximate locations
of the $\ell$-change point using a pilot estimate with large smoothing.
In the second stage, the smoothing is reduced, but we impose the shape
restrictions obtained from the pilot estimate.
Theorems \ref{KZ} and \ref{MMFthm} imply that if the kernel halfwidth 
of the pilot
estimator satisfies $h_N>> \ln[N] N^{-1/(2\ell+3)}$, then spurious
inflection points will occur with a probability smaller than $N^c$
for any $c$. To achieve this result, we use 
an alternate form of the Leadbetter-Cryer integral 
to remove the $N_z o(1)$ from (\ref{CLbnd2}).  


\end{document}